\documentclass{article}
\usepackage[dvips]{graphicx}

\begin{document}
\title{
Enhancement of CP Violating terms \\
for Neutrino Oscillation in Earth Matter}
\author{
\sc
{Keiichi Kimura$^1$}\thanks
{E-mail address:kimukei@eken.phys.nagoya-u.ac.jp} ,
{Akira Takamura$^{1,2}$}\thanks{E-mail
address:takamura@eken.phys.nagoya-u.ac.jp} \\
\sc
{and} \\
\sc
{Hidekazu Yokomakura$^1$}\thanks{E-mail
address:yoko@eken.phys.nagoya-u.ac.jp}
\\
\\
{\small \it $^1$Department of Physics, Nagoya University,}
{\small \it Nagoya, 464-8602, Japan}\\
{\small \it $^2$Department of Mathematics,
Toyota National College of Technology}\\
{\small \it Eisei-cho 2-1, Toyota-shi, 471-8525, Japan}}
\date{}
\maketitle

\vspace{-9.5cm}
\begin{flushright}
  {DPNU 04-13}\\
  {July 2004}
\end{flushright}
\vspace{7.5cm}

\begin{abstract}
We investigate the $\nu_e \to \nu_{\mu}$ oscillation
in the framework of three generations when neutrinos pass through the earth.
The oscillation probability is represented by the form,
$P(\nu_e \to \nu_\mu)=A\cos \delta+B\sin \delta+C$
in arbitrary matter profile by using the leptonic CP phase $\delta$.
We compare our approximate formula in the previous paper
with the formula which includes second order terms of
$\alpha=\Delta m_{21}^2/\Delta m_{31}^2$ and $s_{13}=\sin \theta_{13}$.
Non-perturbative effects of $\alpha$ and
$s_{13}$ can be taken into account in our formula
and the precision of the
formula is rather improved around the MSW resonance region.
Furthermore, we compare the earth matter effect of $A$ and $B$
with that of $C$ studied by other authors.
We show that the magnitude of $A$ and $B$ can reach a few ten \% of $C$
around the main three peaks of $C$ in the region $E>1$ GeV
by numerical calculation.
We give the qualitative understanding of this result by using our
approximate formula.
The mantle-core effect, which is different from the usual MSW
effect, appears not only in $C$ but also in $A$ and $B$,
although the effect is weakened.
\end{abstract}

\section{Introduction}

\hspace*{\parindent}
The first evidence of neutrino oscillation have been discovered
in the atmospheric neutrino experiments and
the mass squared difference $|\Delta m^2_{31}|$ and the 2-3 mixing angle
$\theta_{23}$ \cite{atm} have been measured.
Also the deficit of solar neutrino strongly suggests the neutrino
oscillation
with the Large Mixing Angle (LMA) solution for $\Delta m^2_{21}$
and $\theta_{12}$ \cite{solar}.
This has been confirmed by the KamLAND experiment
by using the artificial neutrino beam emitted from several reactors
\cite{kamland}.
On the other hand, only the upper bound $\sin^2 2\theta_{13} \le 0.1$
is obtained for the 1-3 mixing angle \cite{CHOOZ}.
Thus, the values of the mass differences and the mixing angles are
gradually clarified.
Our aim in the future is to determine the unknown parameters
like the sign of $\Delta m^2_{31}$, $\theta_{13}$ and the leptonic CP phase
$\delta$.

The simple analytic formula for estimating the matter effects
is useful in order to study these parameters
because neutrinos pass through the earth in most experiments
and receive the matter potential represented by $a=\sqrt{2}G_F N_e$,
where $N_e$ is the electron number density and $G_F$ is the Fermi constant.
In the case of short baseline length,
we can approximate the density as constant because
the variation of $N_e$ is small.
However, the longer the baseline is, the larger the matter effect is.
In previous papers, several approximate formulas
have been proposed in order to include the effect of varying density.
Classified by the neutrino energy $E$,
there are following approximate formulas:
low energy formulas by the expansion in the small parameter
$2Ea/|\Delta m_{31}^2| \ll 1$ or $s_{13}=\sin \theta_{13} \ll 1$
\cite{Arafune9703},
high energy formulas by the expansion in
$\Delta m_{21}^2/2Ea \ll 1$ or
$\alpha=\Delta m_{21}^2/\Delta m_{31}^2 \ll1$ \cite{Yasuda9910},
and the formulas by the expansion in
$2E\delta a/\Delta m_{31}^2 \ll 1$ \cite{Ota0011},
where $\delta a$ is the deviation from the average matter potential.

On the other hand, there is the method to approximate
the earth matter density as three constant layers
in the case of mantle-core-mantle \cite{Krastev88}.
It was discussed in refs.
\cite{Ermilova86} how the probability is enhanced
when neutrinos pass through periodically varying density.
Then, it was pointed out in ref. \cite{Petcov9805} that
the mantle-core effect, which is different from the usual
Mikheyev-Smirnov-Wolfenstein (MSW) effect \cite{MSW},
appears in the oscillation probability.
More detailed analysis has been in refs.
\cite{Petcov9903, Akhmedov9805}.
This effect is interesting because the large enhancement of the
probability can occur even if both the effective mixing
angles in the mantle and the core are small.
In recent papers \cite{Akhmedov9808},
the possibility for measuring the $\theta_{13}$ in atmospheric
neutrino experiments has been discussed by using this mantle-core effect.
They concluded that the value of $\theta_{13}$ can be measured
in some cases.

In our previous papers, we have shown that the oscillation probability
for $\nu_e \to \nu_{\mu}$ transition is represented by the following form,
\begin{eqnarray}
\label{eq:kty1}
P(\nu_e \to \nu_\mu)
= \hspace{-1.5em}
 &&   A \cos \delta + B \sin \delta + C
                               \label{1}
\end{eqnarray}
in constant matter \cite{Kimura0203} and also in
arbitrary matter \cite{YKT2}.
By using this general feature for the CP dependence,
the method for solving the parameter ambiguity problem pointed out in
refs. \cite{Ambiguity}
is discussed in ref. \cite{Minakata}.
Each coefficients has an order
$A=O(s_{13}\alpha)$, $B=O(s_{13}\alpha)$ and $C=O(s_{13}^2)+O(\alpha^2)$
on the two small parameters $\alpha=\Delta m_{21}^2/\Delta m_{31}^2\sim
0.04$
and $s_{13}=\sin \theta_{13}< 0.2$.
In the case of $\alpha < s_{13}$,
the ratio of $A$, $B$ to $C$ are given by
$A/C= O(\alpha/s_{13})$ and $B/C= O(\alpha/s_{13})$.
So, it is expected that the CP violating effect due to $A$ and $B$
becomes large and can reach
a few ten \% of $C$ even for the case that neutrinos pass through
the earth core.
However, the effect due to the CP phase has not been taken into account
in previous works.

In this paper,
as the preparation of studying earth matter effect,
we review our approximate formula introduced in ref. \cite{Takamura0403} as
\begin{eqnarray}
A\simeq \hspace{-1.5em} &&
     2c_{23}s_{23}{\rm Re}[{S}_{\mu e}^{\ell *} S_{\tau e}^{h}],
                                \label{2} \\
B\simeq \hspace{-1.5em} &&
     -2c_{23}s_{23}{\rm Im}[{S}_{\mu e}^{\ell *} S_{\tau e}^{h}],
                                \label{3} \\
C\simeq \hspace{-1.5em} &&
|{S}_{\mu e}^{\ell}|^2 c_{23}^2+|S_{\tau e}^{h}|^2 s_{23}^2,
                                \label{4}
\end{eqnarray}
where $S_{\mu e}^{\ell}$ and $S_{\tau e}^h$ are the amplitudes
calculated from two-generation Hamiltonians $H_{\ell}$ and $H_h$.
$H_{\ell}$ is represented by $\Delta m_{21}^2$ and $\theta_{12}$ and
$H_h$ is represented by $\Delta m_{31}^2$ and $\theta_{13}$.
We show that our formula includes the non-perturbative effect of
$\alpha$ and $s_{13}$ and the precision is
rather improved around the MSW resonance regions compared to
the well known simple formula \cite{Cervera0002, Akhmedov0402},
which includes up to second order terms
of $\alpha$ and $s_{13}$.
Furthermore, we compare the earth matter effect of $A$ and $B$
with that of $C$ by using
the Preliminary Reference Earth Model (PREM) \cite{PREM}
in the case of two reference baseline length.
We show that the magnitude of $A$ and $B$ can reach a few ten \% of $C$
around the main three peaks of $C$ in the region $E>1$ GeV
by numerical calculation.
This means that the above perturbative estimation is valid
even in the case of including non-perturbative effect.
We give the qualitative understanding of this result by using our
approximate formula.
The mantle-core effect, which is different from the usual MSW
effect, appears not only in $C$ but also in $A$ and $B$,
although the effect is weakened.

\section{General Formulation for Neutrino Oscillation Probabilities}

\hspace*{\parindent}
In this section, we review the exact formulation of neutrino oscillation
in arbitrary matter profile based on ref. \cite{YKT2}.
At first, let us parametrize the Maki-Nakagawa-Sakata (MNS) matrix $U$
\cite{MNS}, which connects the flavor eigenstate $\nu_\alpha$
with the mass eigenstate $\nu_i$, by the standard parametrization
\cite{PDG}
\begin{eqnarray}
  \label{5}
  U_{}=O_{23}\Gamma_{\delta}O_{13}\Gamma_{\delta}^{\dagger}O_{12},
\end{eqnarray}
where $\Gamma_{\delta}=\mbox{diag}(1,1,e^{i\delta})$ and
$O_{ij}$ is the rotation matrix between $i$ and $j$ generation.
As the matter potential only appears in the ($ee$) component of
the Hamiltonian, $O_{23}$ and $\Gamma_{\delta}$ can be factored out.
So, we can rewrite the Hamiltonian in matter as
$H=O_{23}\Gamma_{\delta} H'(O_{23}\Gamma_{\delta})^\dagger$.
$H'$ is the reduced Hamiltonian defined on the basis
$\nu'=(O_{23}\Gamma_{\delta})^\dagger\nu$.
$H'$ is a real symmetric matrix and the concrete expression is given by
\begin{eqnarray}
{H'}&=&
  O_{13}O_{12}
{\rm diag}(0, \Delta_{21}, \Delta_{31})
 (O_{13}O_{12})^T
+
{\rm diag}(a(t), 0, 0),
 \label{6}
\end{eqnarray}
where $\Delta_{ij}=\Delta m^2_{ij}/2E$.
The number of parameters in the Hamiltonian $H'$ is
fewer than that in the original Hamiltonian $H$ by two.
This is useful to calculate the oscillation probability
simply.
If we define the amplitude $S'_{\alpha\beta}$
of $\nu'_{\alpha} \to \nu'_{\beta}$ transition as
($\alpha\beta$) component of time ordered product
\begin{eqnarray}
  \label{7}
  S'={\rm T} \exp \left[ -i \int^L_0 H'(t)dt \right],
\end{eqnarray}
the oscillation probability is given by
\begin{eqnarray}
P(\nu_e \to \nu_\mu) &=& A \cos\delta
+ B \sin\delta + C, \\
  \label{8}
A&=& 2c_{23}s_{23}{\rm Re}[{S'}_{\mu e}^{*} S'_{\tau e}],
                         \label{9} \\
B&=& -2c_{23}s_{23}
     {\rm Im}[{S'}_{\mu e}^{*} S'_{\tau e}],
                         \label{10} \\
C&=&|{S}'_{\mu e}|^2 c_{23}^2+|S'_{\tau e}|^2 s_{23}^2
                                \label{11}
\end{eqnarray}
as in \cite{YKT2}.
>From the eqns. (\ref{8})-(\ref{11}), one can see that
the probability for $\nu_e \to \nu_{\mu}$ transition
is represented by two components of the reduced amplitude,
$S'_{\mu e}$ and $S'_{\tau e}$.
Namely, the matter effect for the oscillation probability
is only contained in the two components.

\section{Non-Perturbative Effect in Our Approximate Formula}

\hspace*{\parindent}
In this section, we numerically calculate the amplitudes
$S'_{\mu e}$ and $S'_{\tau e}$ introduced in the previous section
by using the PREM.
Then, it is explained how we obtain the hint for the basic concept
on deriving our approximate formula.
As an example, the approximate formula in constant matter
is derived explicitly and is compared with the formula in refs
\cite{Cervera0002, Akhmedov0402}, which includes up to second order
of the small parameters $\alpha=\Delta_{21}/\Delta_{31}$
and $s_{13}$.
As a result, it is shown that our approximate formula includes
the non-perturbative effect which becomes important around
the MSW resonance.

\subsection{Behavior of Reduced Amplitudes in Earth Matter}

\hspace*{\parindent}
Let us calculate the amplitudes $S'_{\mu e}$ and $S'_{\tau e}$
for the case that neutrinos pass through the earth.
We use the PREM as the earth density model and we choose
two reference baselines, $6000$ km and $12000$ km.
Fig. 1 shows how the matter density changes along the path of
neutrinos.
In fig. 2, we plot the values of the amplitudes
$S'_{\mu e}$ and $S'_{\tau e}$
corresponding to the neutrino energy $0.03$-$20$ GeV.
Here, we use the parameters $\Delta m_{21}^2=7\times 10^{-5} {\rm eV}^2$
and $\sin^2 2\theta_{12}=0.8$ as indicated from the solar neutrino
experiments and the KamLAND experiment, $\Delta m_{31}^2=2\times 10^{-3}
{\rm
eV}^2$
from the atmospheric neutrino experiments and the K2K experiment,
and $\sin^2 2\theta_{13}=0.1$ within the upper limit of the CHOOZ
experiment.
\begin{center}
\begin{tabular}{cc}
    $L=6000$ km & $L=12000$ km \\
    \resizebox{86mm}{!}{\includegraphics{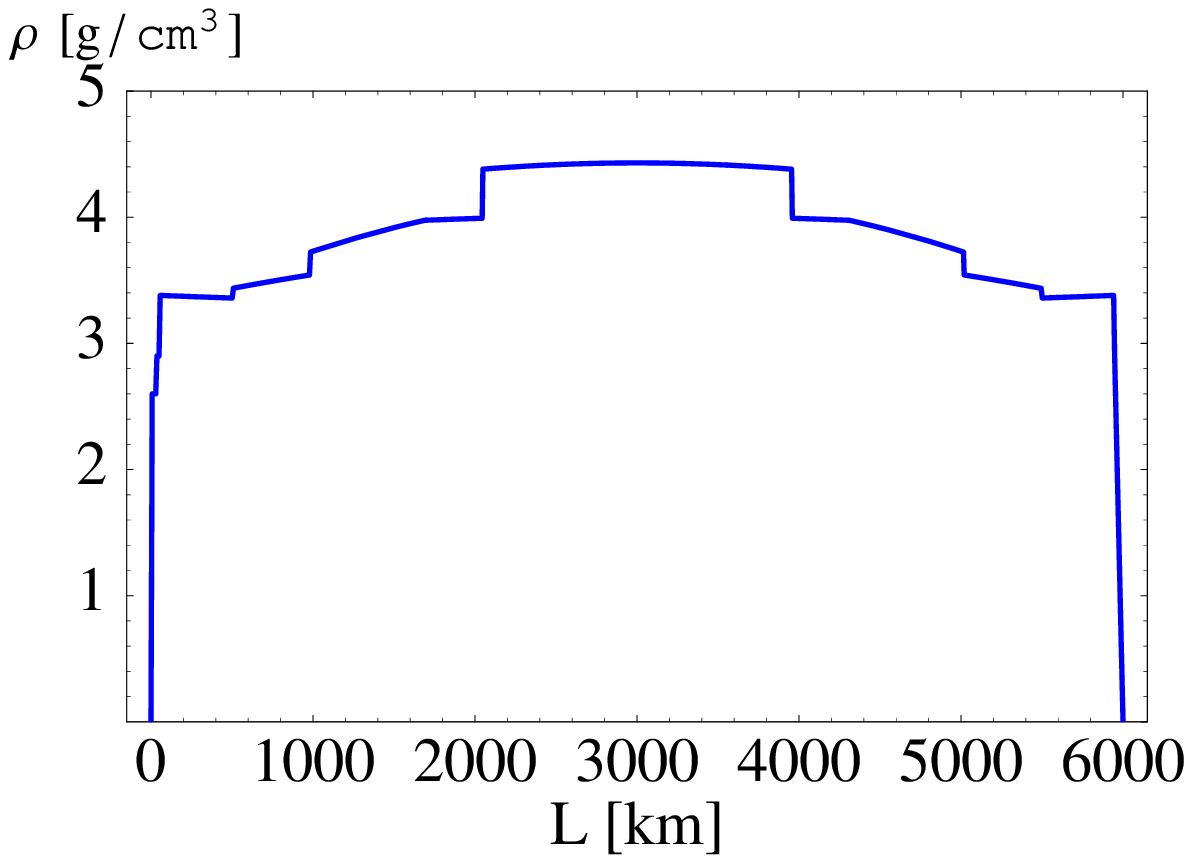}} &
    \resizebox{74mm}{!}{\includegraphics{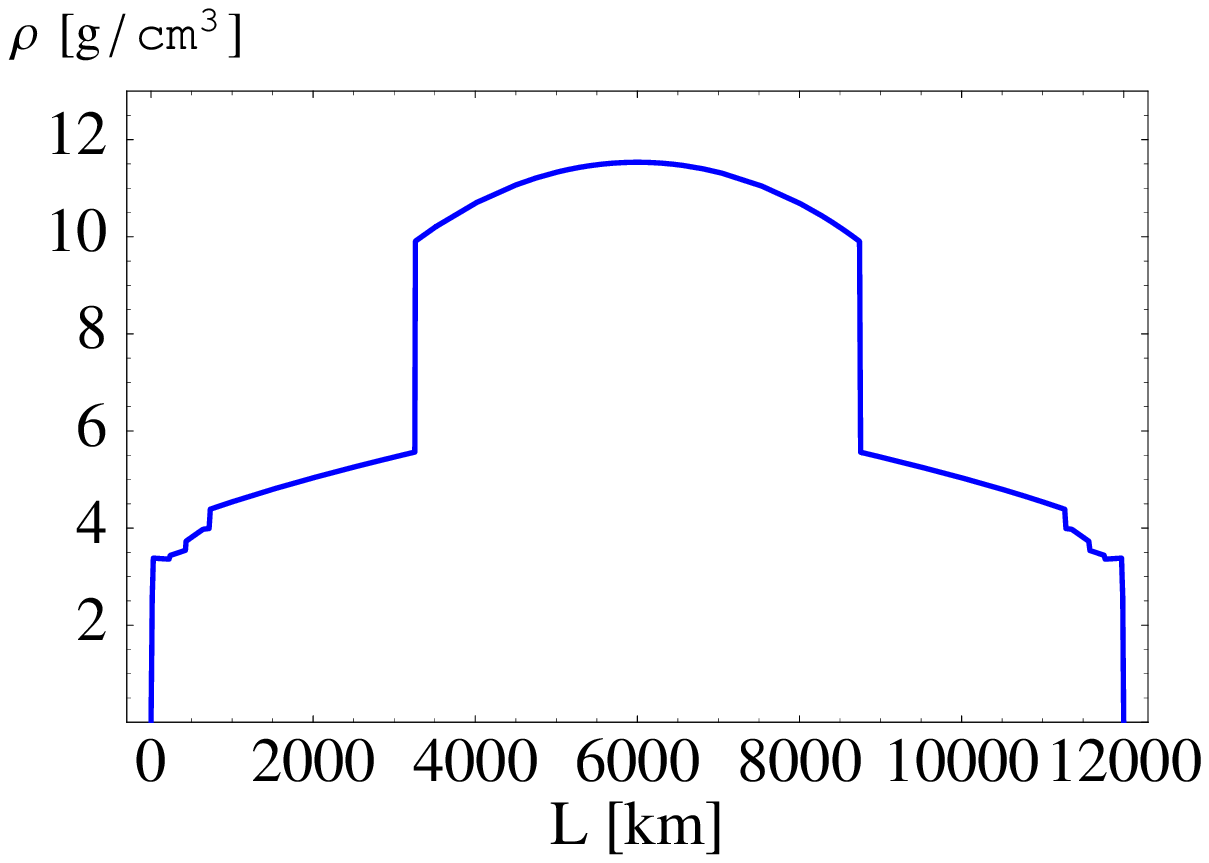}} \\
\end{tabular}

\vspace{-0.2cm}
\begin{flushleft}
Fig. 1. Matter density in the PREM with baseline length
$6000$ km and $12000$ km from left to right.
\end{flushleft}
\end{center}
\begin{center}
\begin{tabular}{cc}
    $L=6000$ km & $L=12000$ km \\
    \resizebox{75mm}{!}{\includegraphics{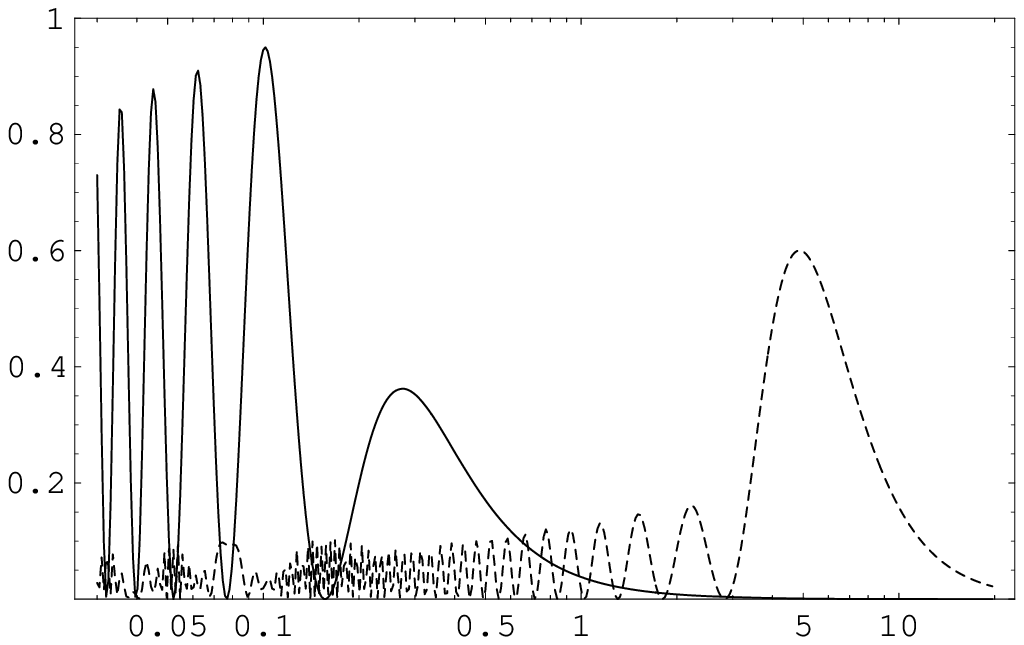}} &
    \resizebox{75mm}{!}{\includegraphics{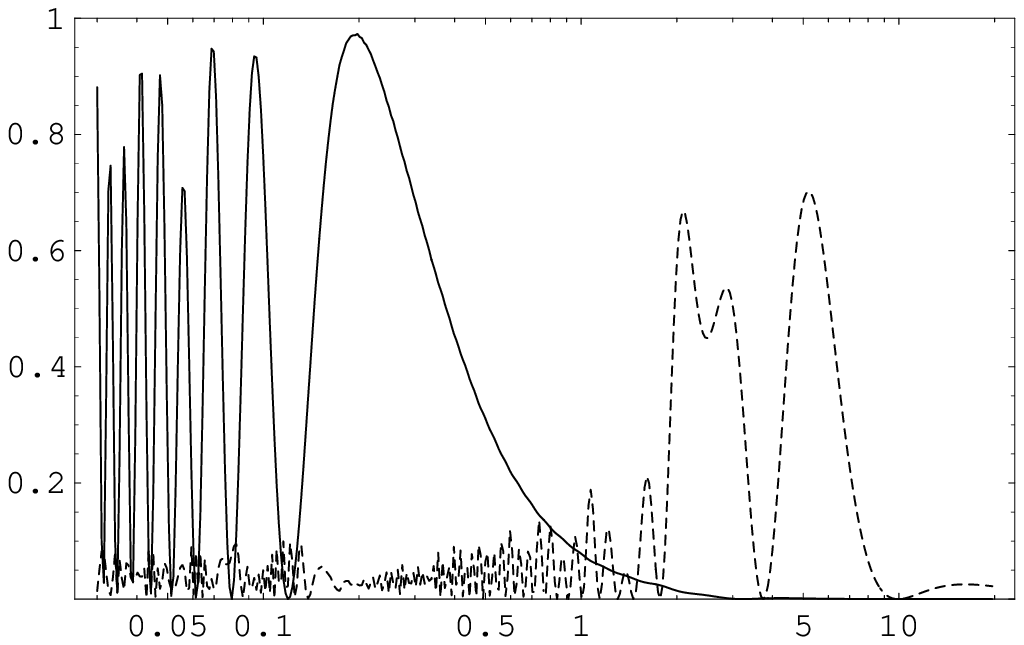}} \\
\qquad E (GeV) & \qquad E (GeV) \\
\end{tabular}
\begin{flushleft}
Fig. 2. Energy dependence of $S'_{\mu e}$ and $S'_{\tau e}$
with baseline length $6000$ km and $12000$ km by using
the PREM. The solid and dashed lines represent
$S'_{\mu e}$ and $S'_{\tau e}$ respectively.
\end{flushleft}
\end{center}

It is found from fig. 2 that $S'_{\mu e}$ and $S'_{\tau e}$
become large in low energy and high energy, respectively,
for both baselines and the regions,
where $S'_{\mu e}$ and $S'_{\tau e}$ dominantly contribute,
are separated to each other.
In other words,
the MSW effect related to the 1-2 mixing and 1-3 mixing angles
are mainly included in $S'_{\mu e}$ and $S'_{\tau e}$, respectively.
We have derived the approximate formula for arbitrary matter profile
by using this feature in ref. \cite{Takamura0403}.
Concretely, $S'_{\mu e}$ and $S'_{\tau e}$ are
calculated from two kinds of different Hamiltonians, which are
given by the 1-2 and 1-3 subsystems, respectively.

\subsection{Procedure of Deriving Approximate Formula}

\hspace*{\parindent}
The idea introduced in the previous subsection is actually realized as
follows.
We use the two small parameters $\alpha=\Delta_{21}/\Delta_{31}$
and $s_{13}$.
Then, our approximate formula is calculated by the following three steps.
\begin{itemize}
\item[1.]
We define two Hamiltonians in the 1-2 and 1-3 subsystems
taking the limit of $s_{13} \to 0$ and $\alpha \to 0$ in (\ref{6}) as
\begin{eqnarray}
  \label{12}
 H_{\ell} =
\left(
     \begin{array}{ccc}
   \Delta_{21} s_{12}^2 + a(t) & \Delta_{21} c_{12} s_{12} & 0 \\
   \Delta_{21} c_{12} s_{12} & \Delta_{21} c_{12}^2 & 0 \\
  0 & 0 & \Delta_{31}
     \end{array}
   \right),
\quad
 H_{h} =
\left(
     \begin{array}{ccc}
   \Delta_{31} s_{13}^2 + a(t) & 0 & \Delta_{31} c_{13} s_{13} \\
   0 & 0 & 0 \\
  \Delta_{31} c_{13} s_{13} & 0 & \Delta_{31} c_{13}^2
     \end{array}
   \right).
\end{eqnarray}

\item[2.]
We calculate two amplitudes $S^{\ell}$ and $S^h$
from the Hamiltonians $H_{\ell}$ and $H_h$ by the equations
\begin{eqnarray}
  \label{13}
  S^{\ell}={\rm T} \exp \left[ -i \int^L_0 H_{\ell}(t)dt \right],
\quad
S^h={\rm T} \exp \left[ -i \int^L_0 H_h(t)dt \right].
\end{eqnarray}

\item[3.]
We replace the amplitudes in (\ref{8})-(\ref{11}) as
$S'_{\mu e} \to S^{\ell}_{\mu e}$ and $S'_{\tau e} \to S^h_{\tau e}$.
\end{itemize}

\subsection{Approximate Formula in Constant Matter}

\hspace*{\parindent}
Next, let us review the approximate formula in constant matter
based on ref. \cite{Takamura0403}.
According to the procedure in the previous subsection,
we substitute the Hamiltonian $H_{\ell}$ in constant matter given by
(\ref{12})
into (\ref{13}) and we obtain $S^{\ell}_{\mu e}$ as
\begin{eqnarray}
  \label{14}
S^{\ell}_{\mu e}=\left[\exp (-iH_{\ell} L) \right]_{\mu e}
=-i\sin 2 \theta_{\ell}
\sin \phi_{\ell} \exp
\left(-i\frac{\Delta_{21}+a}{2}L\right),
\end{eqnarray}
where $\phi_{\ell} \equiv \Delta_{\ell} L/2$ and
the subscript $\ell$ represents the quantities calculated
from $H_{\ell}$.
The concrete expressions for the mass squared difference
and the 1-2 mixing angle in matter are given by
\begin{eqnarray}
\frac{\Delta_{\ell}}{\Delta_{21}}
=\frac{\sin 2\theta_{12}}{\sin 2\theta_{\ell}}
=\sqrt{\left(\cos 2\theta_{12}-\frac{a}{\Delta_{21}}\right)^2
+\sin^2 2\theta_{12}}. \label{15}
\end{eqnarray}
These are well known expressions in the framework of two generations.
The contribution of the low energy MSW effect, which is dominant
around the energy region determined
by $a\sim \Delta_{21}\cos2\theta_{12}$,
is included in mainly
$S_{\mu e}^{\ell}$.
The phase factor in (\ref{14}) does not contribute when we calculate
the probability in two generations.
However, this gives important contribution on the calculation of the terms
dependent on the CP phase in three generations.

Similarly, we obtain $S_{\tau e}^h$ by substituting $H_h$ in constant
matter given
by (\ref{12}) into (\ref{13}) as
\begin{eqnarray}
  \label{16}
&&  S^{h}_{\tau e}=\left[\exp (-iH_h L) \right]_{\tau e}
=-i\sin 2 \theta_{h}
\sin \phi_h
\exp \left(-i\frac{\Delta_{31}+a}{2}L\right),
\end{eqnarray}
where $\phi_h \equiv \Delta_{h} L/2$ and
the subscript $h$ represents the quantities calculated
from $H_h$.
The concrete expressions are given by
\begin{eqnarray}
\frac{\Delta_{h}}{\Delta_{31}}
=\frac{\sin 2\theta_{13}}{\sin 2\theta_{h}}
=\sqrt{\left(\cos 2\theta_{13}-\frac{a}{\Delta_{31}}\right)^2
+\sin^2 2\theta_{13}}. \label{17}
\end{eqnarray}
One can see that these expressions correspond to those obtained
by the replacement $\Delta_{21}\to\Delta_{31}$ and
$\theta_{12}\to \theta_{13}$ in (\ref{14}) and (\ref{15}).
The contribution of high energy MSW effect, which is dominant
around the energy region determined by
$a\sim \Delta_{31}\cos 2\theta_{13}$, is included
in mainly $S_{\tau e}^h$.
We can calculate $A$, $B$ and $C$ in constant matter as
\begin{eqnarray}
P(\nu_e \to \nu_{\mu})=A\cos \delta+B\sin \delta+C,
\label{18}
\end{eqnarray}
\begin{eqnarray}
A &\simeq& \sin 2\theta_{\ell}\sin 2\theta_{23}\sin 2\theta_{h}
\sin \phi_{\ell} \sin \phi_h
\cos \frac{\Delta_{32}L}{2},
\label{19} \\
B &\simeq& \sin 2\theta_{\ell}\sin 2\theta_{23}\sin 2\theta_{h}
\sin \phi_{\ell} \sin \phi_h
\sin \frac{\Delta_{32}L}{2},
\label{20} \\
C &\simeq&
 c_{23}^2\sin^2 2\theta_{\ell}\sin^2 \phi_{\ell}
+ s_{23}^2\sin^2 2\theta_{h}\sin^2 \phi_h
\label{21}
\end{eqnarray}
from these expressions.
These approximate formulas are similar to the following
well known formulas
\begin{eqnarray}
A &\simeq&
\frac{\Delta_{21}\Delta_{31}}{a(a-\Delta_{31})}
c_{13}\sin 2\theta_{12}\sin 2\theta_{23}\sin 2\theta_{13}
\sin \frac{aL}{2}
\sin \frac{(a-\Delta_{31})L}{2}
\cos \frac{\Delta_{31}L}{2},
\label{22} \\
B &\simeq&
\frac{\Delta_{21}\Delta_{31}}{a(a-\Delta_{31})}
c_{13}\sin 2\theta_{12}\sin 2\theta_{23}\sin 2\theta_{13}
\sin \frac{aL}{2}
\sin \frac{(a-\Delta_{31})L}{2}
\sin \frac{\Delta_{31}L}{2},
\label{23} \\
C &\simeq& \frac{\Delta_{21}^2}{a^2}
 c_{23}^2\sin^2 2\theta_{12}
\sin^2 \frac{aL}{2}
+ \frac{\Delta_{31}^2}{(a-\Delta_{31})^2}s_{23}^2\sin^2 2\theta_{13}
\sin^2 \frac{(a-\Delta_{31})L}{2}.
\label{24}
\end{eqnarray}
These formulas are often used in order to analyze the property of
neutrino oscillation because they have very simple form and
approximate the exact values with a good precision.

In the following, we compare the probability calculated from
our approximate formula (\ref{19})-(\ref{21}) with
that from the formula (\ref{22})-(\ref{24})
in the case of constant matter.
We calculate $P(\nu_e \to \nu_{\mu})$ for two kinds of baselines,
$3000$ km and $6000$ km.
We use the parameters $\sin^2 2\theta_{23}=1$ and $\delta=0^{\circ}$
in addition to those introduced in fig. 2.
Furthermore, $\rho=4.7$g/cm$^3$ and $Y_e=0.494$ are used as
the matter density and the electron fraction.
The result is given in fig. 3.

\begin{center}
\begin{tabular}{cc}
    $L=3000$ km & $L=6000$ km \\
    \resizebox{75mm}{!}{\includegraphics{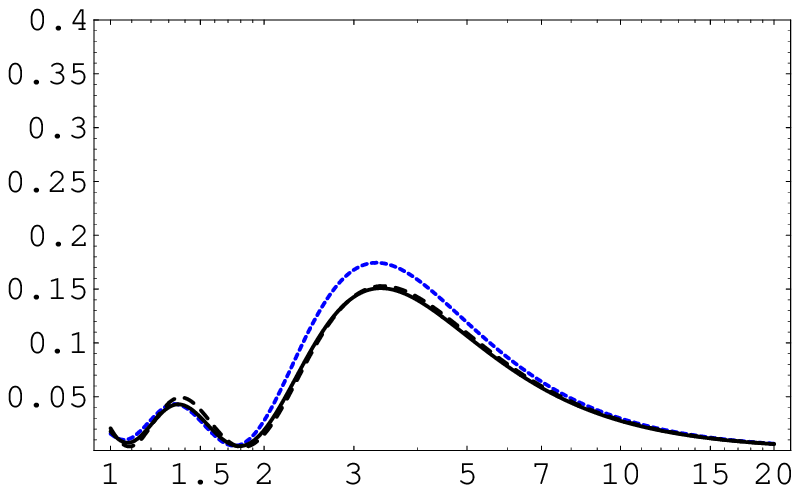}} &
    \resizebox{75mm}{!}{\includegraphics{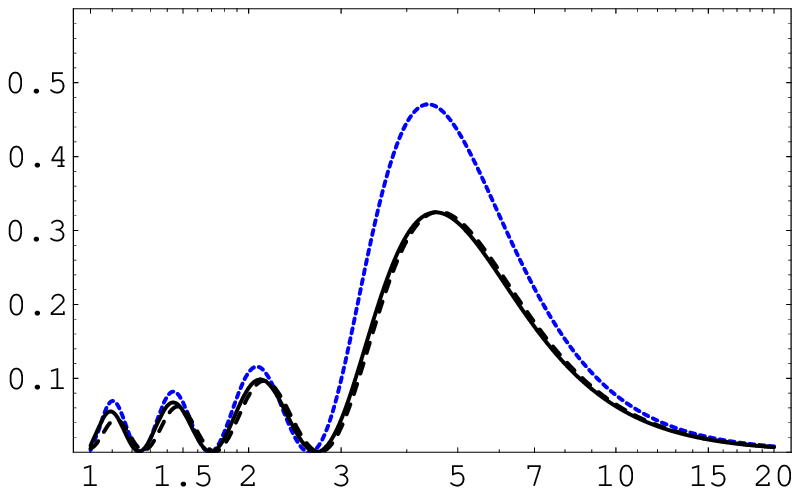}} \\
\qquad E (GeV) & \qquad E (GeV) \\
\end{tabular}
\begin{flushleft}
Fig. 3. Comparison between the probabilities $P(\nu_e \to \nu_{\mu})$
 calculated from our approximate formula (\ref{19})-(\ref{21}) and
the formula (\ref{22})-(\ref{24}).
Two baseline length are chosen as $3000$ km and $6000$ km
from left to right.
The solid, dashed and dotted lines are the exact, our approximate
 formula
and that from (\ref{22})-(\ref{24}).
\end{flushleft}
\end{center}

It is found that our approximate formula has good coincidence
to the exact one even around the MSW resonance, compared with
that from the formula (\ref{22})-(\ref{24}).
We consider the reason for the difference in the next section.

\subsection{Comparison of Approximate Formulas}

\hspace*{\parindent}
Let us explain the order counting of $\alpha$ and $s_{13}$
in our approximate formula.
In the limit $\alpha \to 0$, we obtain $S'_{\mu e}=0$.
Therefore, we can write down the order of $S'_{\mu e}$ as
\begin{eqnarray}
S'_{\mu e}&=&O(\alpha)+O(\alpha^2)+O(\alpha^3)+\cdots \nonumber \\
&+&O(s_{13}\alpha)+O(s_{13}\alpha^2)+O(s_{13}\alpha^3)+\cdots \nonumber \\
&+&O(s_{13}^2\alpha)+O(s_{13}^2\alpha^2)+O(s_{13}^2\alpha^3)+\cdots
\nonumber \\
&+&\cdots\cdots.
\label{25}
\end{eqnarray}
Note that $\alpha$ is included in all terms.
Here, if we take the $s_{13} \to 0$, only the first line is remaining.
In these terms, all orders of $\alpha$ are included and
the first line considered to have a larger contribution
compared to the following
lines, because of the increasing exponent of $s_{13}$.
This is confirmed by the comparison with the exact formula in fig. 3.
In the same way, we obtain $S'_{\tau e}=0$ in the limit
$s_{13} \to 0$.
Therefore, we can write down the order of $S'_{\tau e}$ as
\begin{eqnarray}
S'_{\tau e}&=&O(s_{13})+O(s_{13}^2)+O(s_{13}^3)+\cdots \nonumber \\
&+&O(s_{13}\alpha)+O(s_{13}^2\alpha)+O(s_{13}^3\alpha)+\cdots \nonumber \\
&+&O(s_{13}\alpha^2)+O(s_{13}^2\alpha^2)+O(s_{13}^3\alpha^2)+\cdots
\nonumber \\
&+&\cdots\cdots.
\label{26}
\end{eqnarray}
Here, if we take the limit $\alpha \to 0$, only the first line is remaining.
In these terms, all orders of $s_{13}$ are included and
the first line is considered to have a larger contribution
compared to the following lines,
because of the increasing exponent of $\alpha$.

Our method includes both, the terms of higher order of $\alpha$ in
(\ref{25}) and also those of $s_{13}$ in (\ref{26}).
So, this new approach is not a systematic expansion.
However, our method is not in contradiction to the well known formula
(\ref{22})-(\ref{24}),
which takes only the first order term of $\alpha$ and $s_{13}$
in (\ref{25})-(\ref{26}) regarding them as small parameters.
In addition, higher order terms of the perturbative expansion,
which are not included in the formula (\ref{22})-(\ref{24}),
are now also included in our formula.

In the following, let us investigate the difference
between these two methods more concretely.
As seen in fig. 2, the contribution of $S'_{\tau e}$ is dominant
in the energy region $E>1$ GeV.
So, we can roughly consider as
\begin{eqnarray}
P(\nu_e \to \nu_{\mu})\simeq C \simeq
s_{23}^2\sin^2 2\theta_{h}\sin^2
\left(\frac{\Delta_{h}L}{2}\right).
\label{27}
\end{eqnarray}
Note that we have the relation (\ref{17})
between the mass squared differences and
the mixing angles in vacuum and in matter.
Expanding the right hand side of (\ref{17})
on the mixing angle in vacuum, we obtain
\begin{eqnarray}
\frac{\Delta_{h}}{\Delta_{31}}
=\frac{\sin 2\theta_{13}}{\sin 2\theta_{h}}
\simeq \left|1-\frac{a}{\Delta_{31}}\right|
\left(1+\frac{2a\Delta_{31}}{(\Delta_{31}-a)^2}
s_{13}^2 + \frac{a^2 \Delta_{31}^2}{2(\Delta_{31}-a)^4}s_{13}^4+
\cdots \right).
\label{28}
\end{eqnarray}
The condition for convergence is given by
\begin{eqnarray}
\frac{4a\Delta_{31} s_{13}^2}{(\Delta_{31}-a)^2}<1.
\label{29}
\end{eqnarray}
This condition is not satisfied around the MSW resonance region,
where $\Delta_{31} \sim a$.
Namely, the perturbative expansion becomes inconvergent.
However, substituting the above expression (\ref{28}) into (\ref{27})
of the oscillation probability and taking only the first
term, we obtain
\begin{eqnarray}
P(\nu_e \to \nu_{\mu})\simeq s_{23}^2
\frac{\Delta_{31}^2 \sin^2 2\theta_{13}}{(\Delta_{31}-a)^2}
\sin^2 \frac{\Delta_{31}-a}{2}L.
\label{30}
\end{eqnarray}
It gives the finite value in the limit $\Delta_{31} \to a$ as
\begin{eqnarray}
P(\nu_e \to \nu_{\mu})\simeq s_{23}^2
c^2_{13} (s_{13} \Delta_{31} L)^2.
\label{31}
\end{eqnarray}
This is due to the product of the infinity of effective mixing angle
and the zero of the effective mass squared difference
in the probability (\ref{30}).
If we take the limit $\Delta_{31} \to a$ directly in the
non-perturbative expression (\ref{27}),
we obtain
\begin{eqnarray}
P(\nu_e \to \nu_{\mu})=s_{23}^2 c^2_{13}
\sin^2 (s_{13} \Delta_{31} L).
\label{32}
\end{eqnarray}
We find that the difference between the perturbative formula
(\ref{31}) and our formula (\ref{32}) is the $\sin$ factor.
In the case of the short baseline length $L$, the perturbation
gives a good approximation, but the longer the baseline $L$ is,
the worse the perturbation becomes, as shown in fig. 3,
although the probability has a finite value.
Concretely, if the condition
\begin{eqnarray}
L<\frac{1}{s_{13} \Delta_{31}}
\label{33}
\end{eqnarray}
is satisfied, the perturbation gives a good approximation.
Around the MSW resonance region, the perturbation breaks down
because the coefficients of the higher order terms
$\alpha$ or $s_{13}$ become large.
Therefore, it is needed to involve the higher order terms
of $\alpha$ and $s_{13}$, in order to make a good
approximate formula.
Our method partially realizes this request.

At the end of this section, let us give a brief comment.
In ref. \cite{Cervera0002, Akhmedov0402}, the
formula calculated by single expansion on $\alpha$
is also given and this includes all order terms
of $s_{13}$.
So, this approximate formula gives a good approximation
in the high energy MSW resonance region compared with the formula
(\ref{22})-(\ref{24}), while the difference between
the single expansion formula and numerical calculation becomes large in
the low energy region as commented also in ref. \cite{Akhmedov0402}.

\section{Earth Matter Effect for $A$, $B$ and $C$}

\hspace*{\parindent}
In this section, we perform numerical calculations of $A$, $B$ and $C$
by using the PREM.
We give a qualitative understanding of the behavior of
the coefficients $A$, $B$ and $C$
by matter effects of the mantle and the core.

\subsection{Numerical Calculation of $A$, $B$ and $C$}

\hspace*{\parindent}
In the previous section, the order of the reduced amplitudes are
estimated as $S'_{\mu e}=O(\alpha)$ and $S'_{\tau e}=O(s_{13})$
in the case that we takes only the first order term of
$\alpha$ and $s_{13}$ in (\ref{25})-(\ref{26}).
The order of coefficients are also obtained as $A=O(s_{13}\alpha)$,
$B=O(s_{13}\alpha)$ and $C=O(s_{13}^2)+O(\alpha^2)$
by substituting $S'_{\mu e}=O(\alpha)$ and $S'_{\tau e}=O(s_{13})$
into (\ref{11})-(\ref{13}).
In the case of $\alpha < s_{13}$,
the magnitude of ratios is give by $A/C= O(\alpha/s_{13})$ and $B/C=
O(\alpha/s_{13})$.
Therefore, it is expected from the perturbative point of view
that the CP violating effect due to $A$ and $B$
becomes large and can reach a few ten \% of $C$.
However, because non-perturbative effect becomes important in MSW region
as shown in fig. 3, the CP violating effect should be
investigated more carefully.

At first, let us numerically calculate how the coefficients $A$, $B$
and $C$ are enhanced by the earth matter effect,
in the case of the baseline length $L=6000$km and $12000$km
for $\sin^2 2\theta_{13}=0.10$, and $0.04$, respectively.
These values of $\sin^2 2\theta_{13}$ correspond to the
values within the upper bound of the CHOOZ experiments.
The PREM is used as the earth matter density and
the same mass squared differences and the mixing angles given in sec. 3
are also used.

\begin{center}
\begin{tabular}{ccc}
A ($\sin^2 2\theta_{13}=0.10$, $6000$ km) &
B ($\sin^2 2\theta_{13}=0.10$, $6000$ km) &
C ($\sin^2 2\theta_{13}=0.10$, $6000$ km) \\
    \resizebox{50mm}{!}{\includegraphics{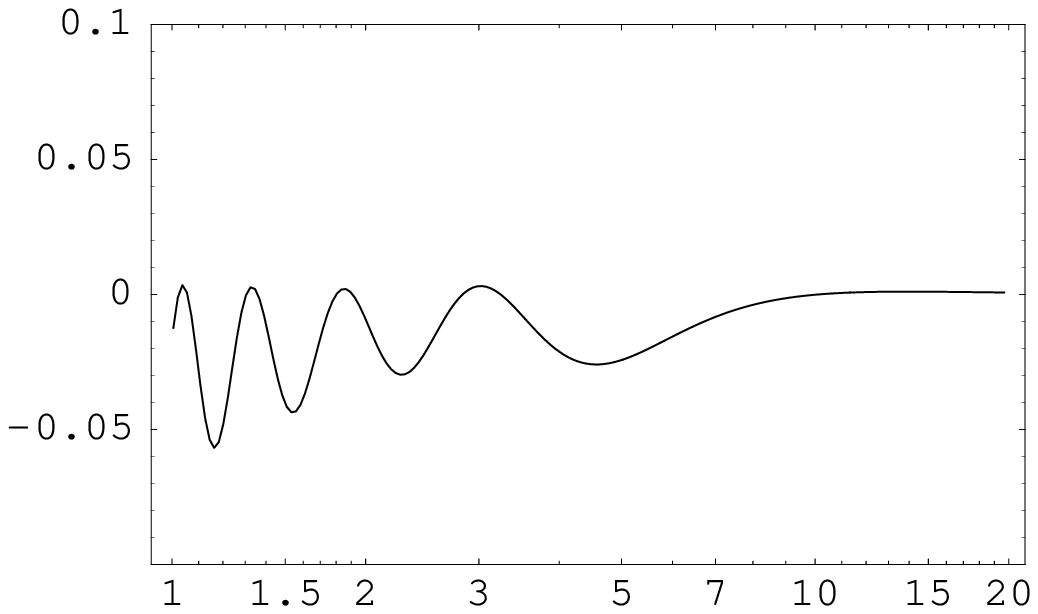}} &
    \resizebox{50mm}{!}{\includegraphics{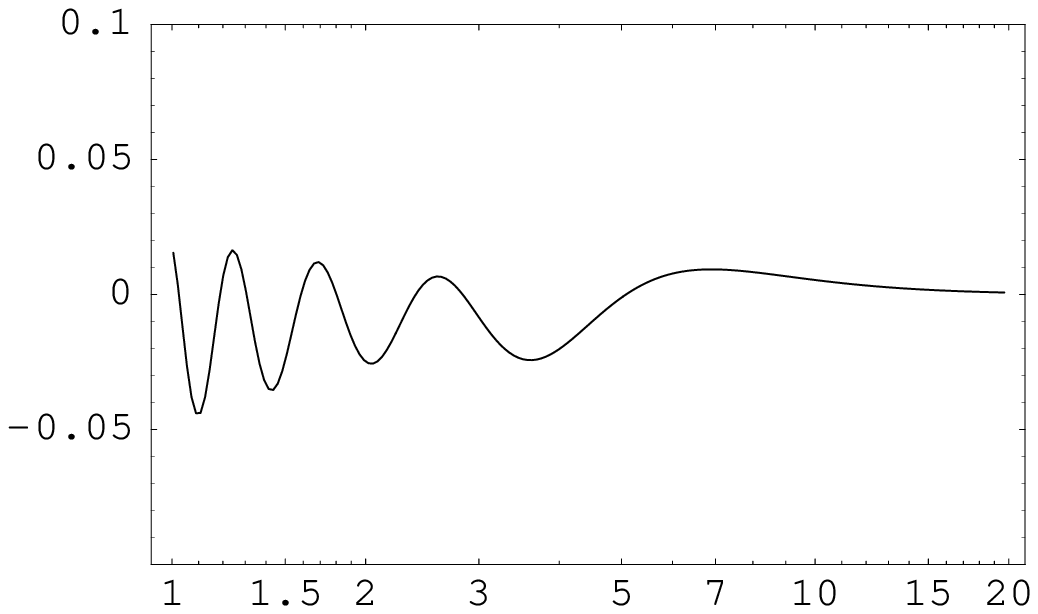}} &
    \resizebox{50mm}{!}{\includegraphics{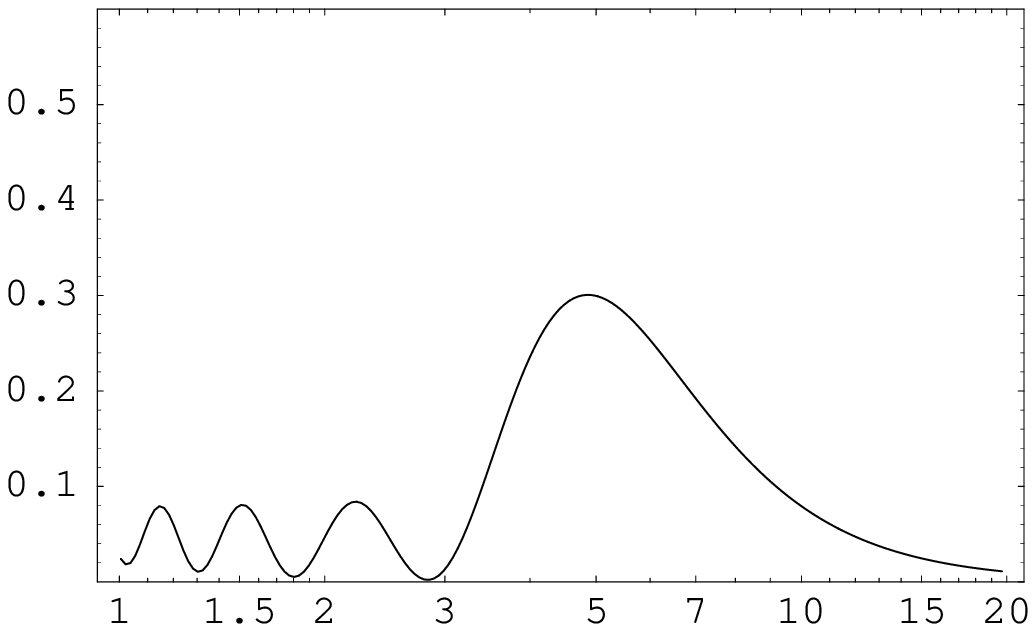}} \\
\qquad E (GeV) & \qquad E (GeV) & \qquad E (GeV) \\
A ($\sin^2 2\theta_{13}=0.04$, $6000$ km) &
B ($\sin^2 2\theta_{13}=0.04$, $6000$ km) &
C ($\sin^2 2\theta_{13}=0.04$, $6000$ km) \\
    \resizebox{50mm}{!}{\includegraphics{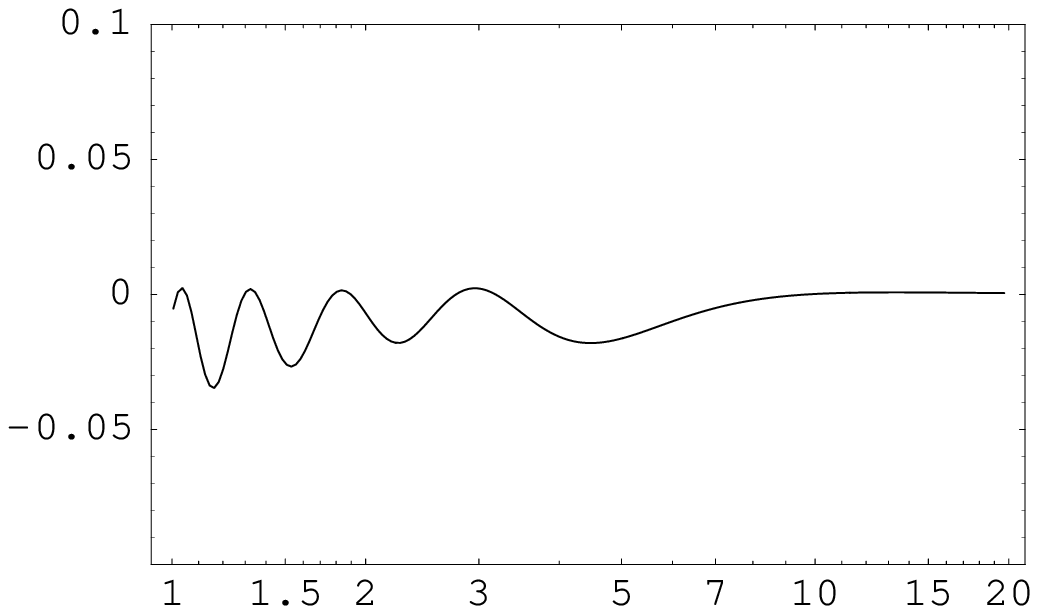}} &
    \resizebox{50mm}{!}{\includegraphics{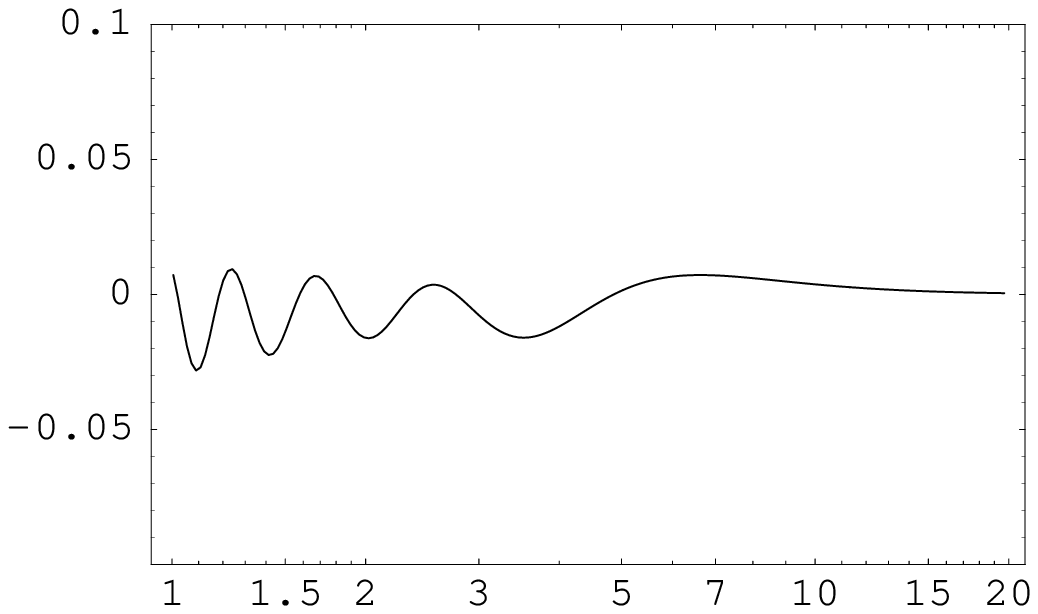}} &
    \resizebox{50mm}{!}{\includegraphics{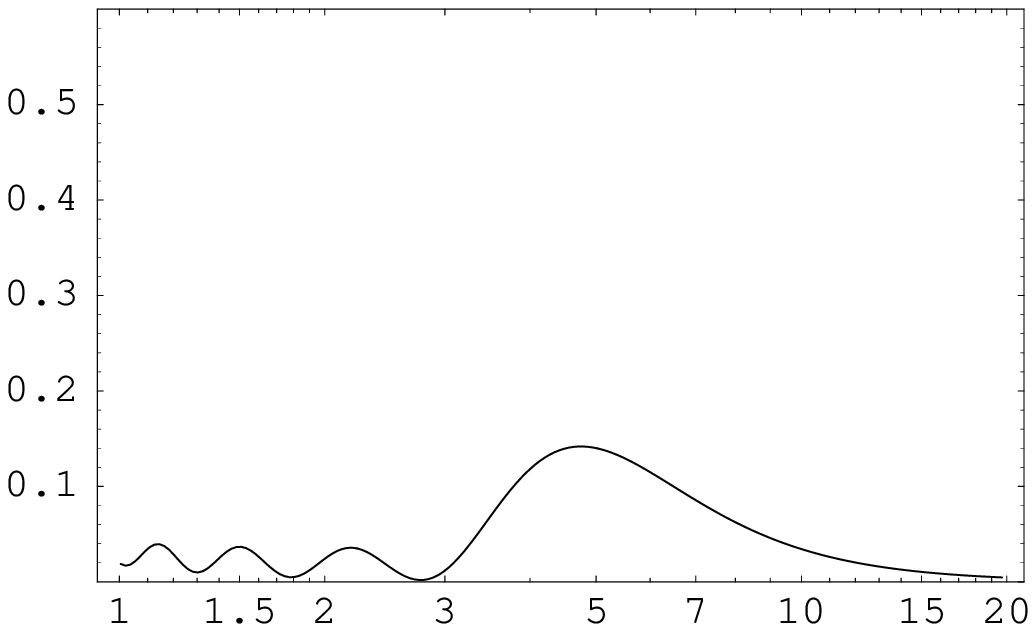}} \\
\qquad E (GeV) & \qquad E (GeV) & \qquad E (GeV) \\
A ($\sin^2 2\theta_{13}=0.10$, $12000$ km) &
B ($\sin^2 2\theta_{13}=0.10$, $12000$ km) &
C ($\sin^2 2\theta_{13}=0.10$, $12000$ km) \\
    \resizebox{50mm}{!}{\includegraphics{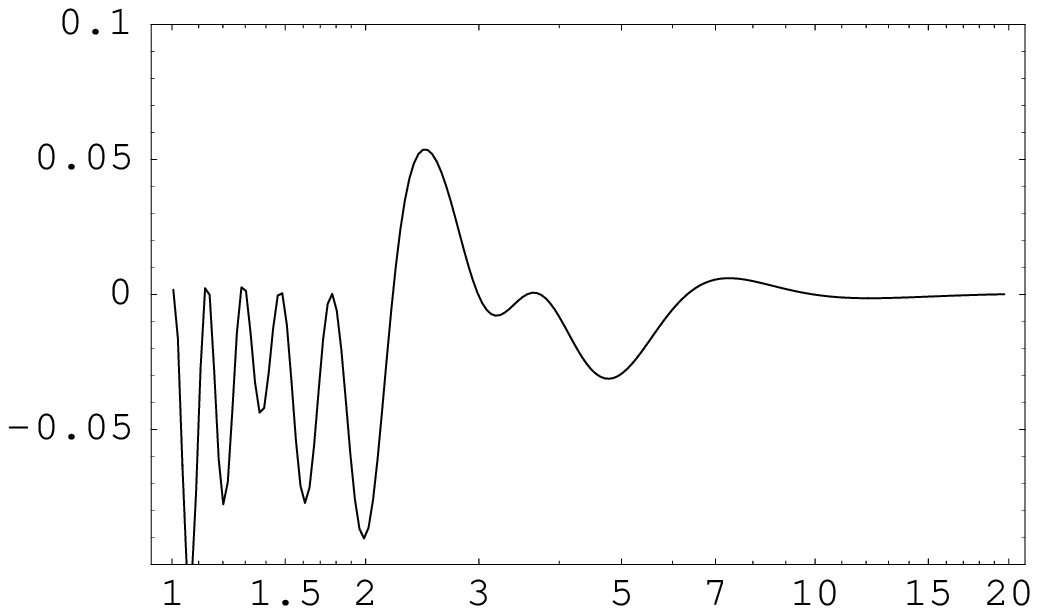}} &
    \resizebox{50mm}{!}{\includegraphics{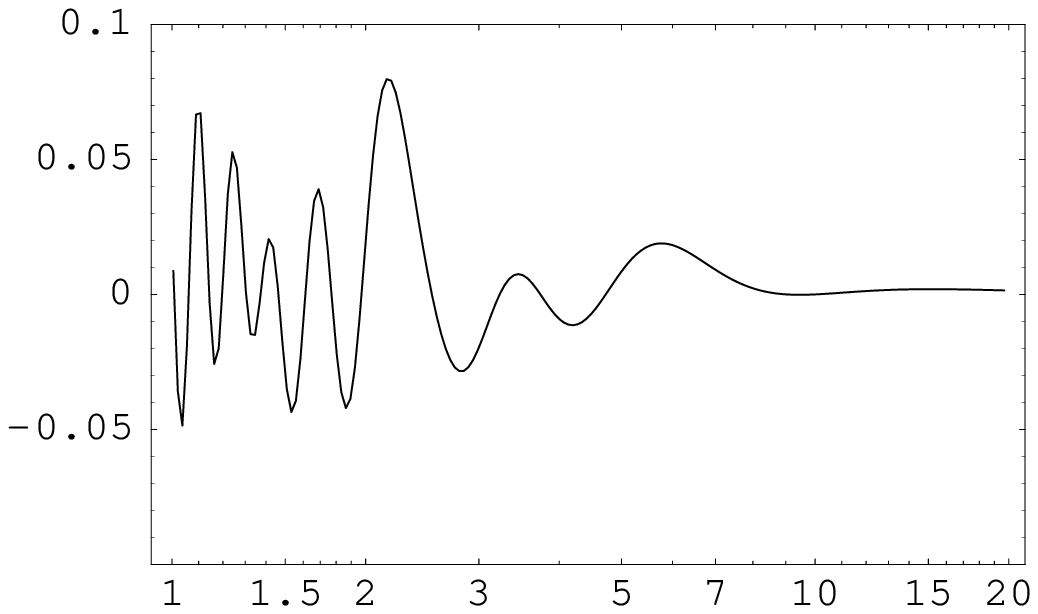}} &
    \resizebox{50mm}{!}{\includegraphics{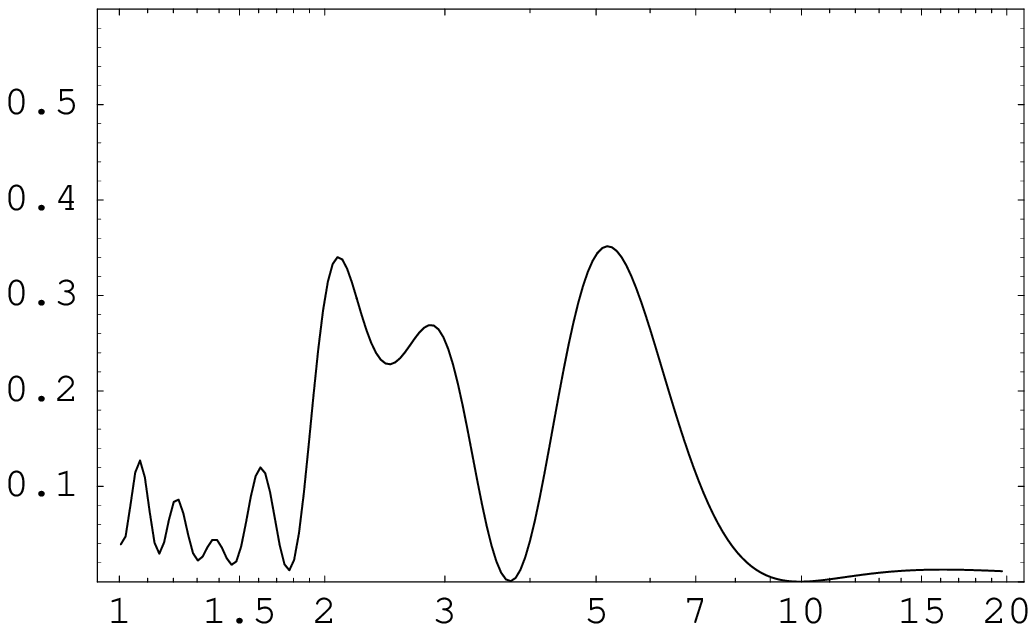}} \\
\qquad E (GeV) & \qquad E (GeV) & \qquad E (GeV) \\
A ($\sin^2 2\theta_{13}=0.04$, $12000$ km) &
B ($\sin^2 2\theta_{13}=0.04$, $12000$ km) &
C ($\sin^2 2\theta_{13}=0.04$, $12000$ km) \\
    \resizebox{50mm}{!}{\includegraphics{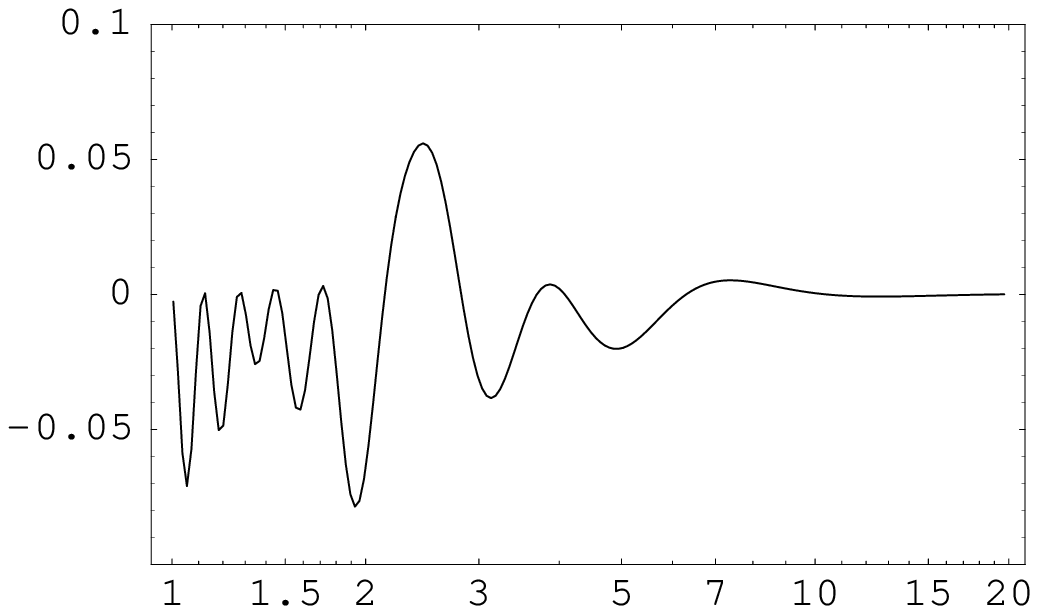}} &
    \resizebox{50mm}{!}{\includegraphics{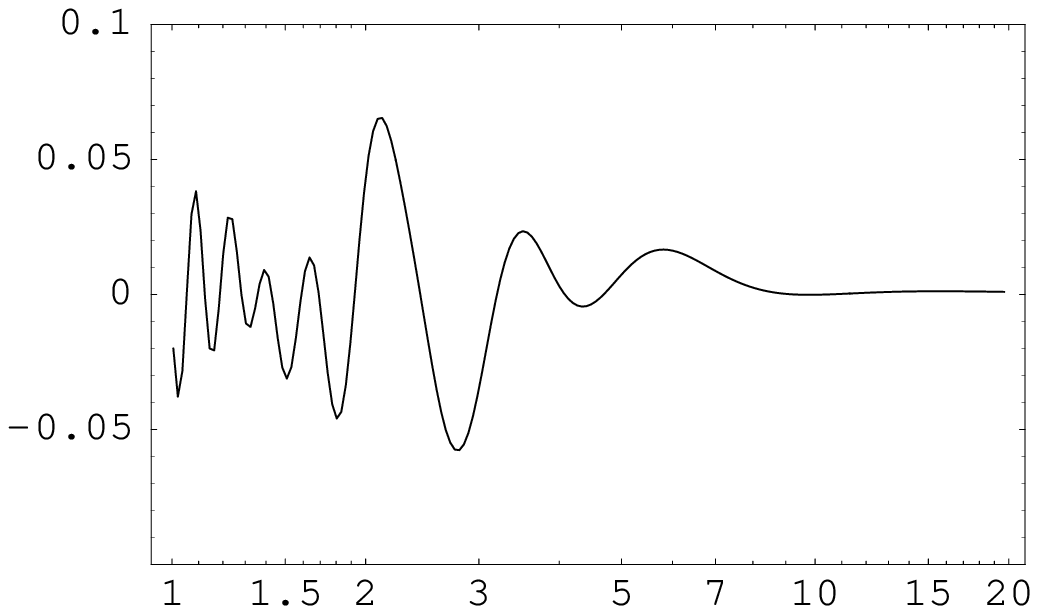}} &
    \resizebox{50mm}{!}{\includegraphics{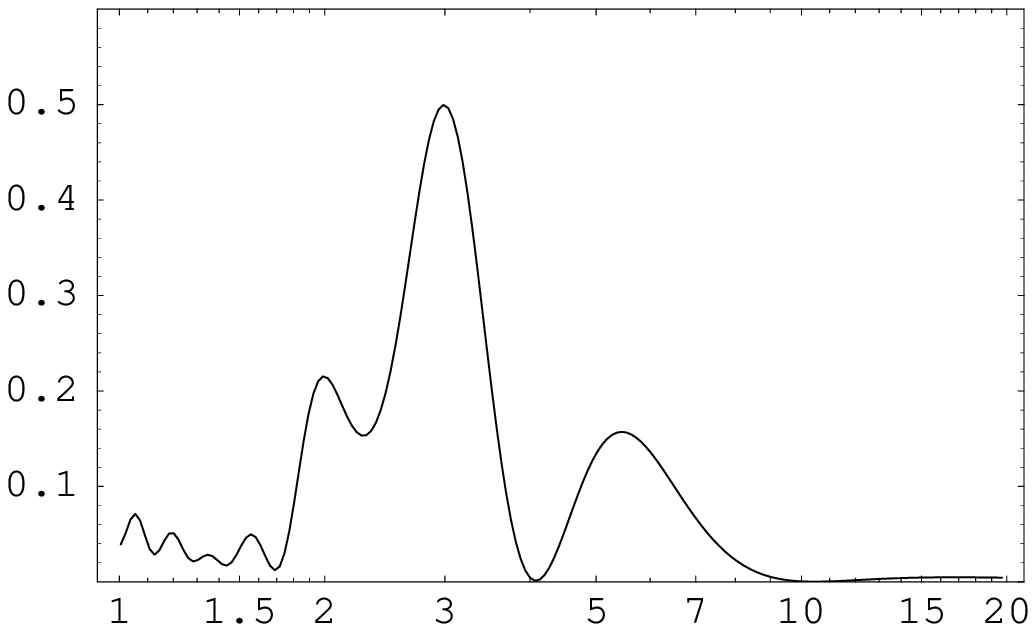}} \\
\qquad E (GeV) & \qquad E (GeV) & \qquad E (GeV) \\
\end{tabular}
\begin{flushleft}
Fig. 4. Energy dependence of $A$, $B$ and $C$ by numerical calculation.
We use the PREM as earth matter density model
with two baseline length $6000$ km and $12000$ km,
and we choose the 1-3 mixing angle $\sin^2 2\theta_{13}=0.10$ and $0.04$
as representative values.
\end{flushleft}
\end{center}

In the case of $L=6000$ km, the behavior can be understood
by using the formulation (\ref{19})-(\ref{21}) in constant
matter.
The value of $C$ becomes large around $E=5$ GeV,
which comes from the enhancement of the effective mixing
angle $\sin \theta_h$ for $a\simeq \Delta_h$,
and then oscillates depending on the factor $\sin \phi_h$.
The values of $A$ and $B$ become small compared with that of
$C$ in high energy region,
as the suppression factor $S^{\ell}_{\mu e}\propto 1/E$.
See details in refs. \cite{Kimura0203, Cervera0002, Harrison9912}
for example of the constant matter density.

In the case of $L=12000$ km,
three main peaks appear in $C$.
On the other hand,
the pattern of the enhancement for $A$ and $B$ seems to become
more complicated than that of $L=6000$ km.
In the next subsection, we give a qualitative understanding of the above
results
by using our approximate formula for $A,B$ and $C$.

We also represent the values of $A$, $B$ and $C$
around the energy of the three peaks of $C$ in Table 1.
These values are computed by the numerical calculation using the PREM.
\begin{center}
\begin{minipage}{14cm}
\begin{tabular}{cccccccc} \hline
$L$ (km) & $\sin^2 2\theta_{13}$ & $E$ (GeV)
& $A$\quad($|A/C|$) & $B$\quad($|B/C|$) & $C$ & peak type
 \\ \hline
$6000$ & $0.10$ & $4.9$ & $-0.025$\quad(8.3\%)
& $-0.003$\quad(1.0\%) & $0.301$ & MSW(mantle) \\
$6000$ & $0.04$ & $4.8$ & $-0.017$\quad(12.0\%)
& $-0.001$\quad(0.7\%) & $0.142$ & MSW(mantle) \\
\hline
$12000$ & $0.10$ & $2.1$ & $-0.061$\quad(17.9\%)
& $0.066$\quad(19.4\%) & $0.340$ & MSW(core) \\
$12000$ & $0.10$ & $2.8$ & $0.017$\quad(6.3\%)
& $-0.028$\quad(10.4\%) & $0.269$ & mantle-core \\
$12000$ & $0.10$ & $5.2$ & $-0.026$\quad(7.4\%)
& $0.013$\quad(3.7\%) & $0.352$ & MSW(mantle) \\
\hline
$12000$ & $0.04$ & $2.0$ & $-0.068$\quad(31.6\%)
& $0.038$\quad(17.7\%) & $0.215$ & MSW(core) \\
$12000$ & $0.04$ & $3.0$ & $-0.030$\quad(6.0\%)
& $-0.037$\quad(7.4\%) & $0.500$ & mantle-core \\
$12000$ & $0.04$ & $5.4$ & $-0.014$\quad(8.9\%)
& $0.015$\quad(9.6\%) & $0.157$ & MSW(mantle) \\
\hline
\end{tabular}
\end{minipage}
\end{center}
\begin{flushleft}
Table 1. Resonance values of $A$, $B$ and $C$ calculated numerically
by using the PREM with baseline length $6000$ km and $12000$ km,
$\sin^2 2\theta_{13}=0.10$ and $0.04$.
\end{flushleft}

Table 1 shows that the coefficients $A$ and $B$ can be rather large
at the three main peaks of $C$.
The absolute values of $A$ and $B$ become about $0.06$ at the peak
of the core.
Furthermore, the ratios $|A/C|$ and $|B/C|$ also
become a few ten \% even for the case of including non-perturbative
effect.
These energy regions $E=2\sim 6$ GeV are explored by the
atmospheric neutrino and the long baseline experiments.
In actual experiments averaging of various parameters are necessary
for example, energy, zenith-angle distribution,
the sum of particle and anti-particle, and so on.
Therefore, the CP phase effect may be weakened to some extent,
but we consider that the CP phase effect should be estimated
precisely in order to determine the value of $\theta_{13}$ in future
experiments.

\subsection{Approximate Formula in Matter with Three Layers}

\hspace*{\parindent}
In order to give a qualitative understanding of the results
obtained in the previous subsection,
let us approximate the earth matter density with baseline
$L=12000$ km as three constant layers such that the first and the third
layers have the same density and length.

At first, we calculate the amplitude $S_{\mu e}^{\ell}$
from the low energy Hamiltonian $H_{\ell}$.
We use the superscript $m$ and $c$ for representing the amplitude
in the first and third layer (mantle), and the second layer
(core).
Taking the limit $s_{13} \to 0$, the amplitudes $S_{\tau e}^m$,
$S_{\tau\mu}^m$ and so on vanish.
Only four terms contribute to
the amplitude in three layers $S_{\mu e}^{\ell}$ as
\begin{eqnarray}
S_{\mu e}^{\ell}
&=& S^m_{\mu e}S^c_{ee}S^m_{ee}
+ S^m_{\mu\mu}S^c_{\mu e}S^m_{ee}
+ S^m_{\mu e}S^c_{e\mu}S^m_{\mu e}
+ S^m_{\mu\mu}S^c_{\mu\mu}S^m_{\mu e}.
\label{34}
\end{eqnarray}
Substituting (\ref{14}) and
\begin{eqnarray}
  \label{35}
&&  S^m_{ee}
=\left[\exp (-iH_{\ell}^m L) \right]_{ee}
=\left(
\cos \phi_{\ell}^m+i\cos 2 \theta^{m}_{\ell}
\sin \phi^m_{\ell} \right)
\exp \left(-i\frac{\Delta_{21}+a^m}{2}L^m\right),
\\
&&  S^m_{\mu\mu}=\left[\exp (-iH_{\ell}^m L) \right]_{\mu \mu}
=\left(\cos \phi_{\ell}^m-i\cos 2 \theta^{m}_{\ell}
\sin \phi_{\ell}^m\right)
\exp \left(-i\frac{\Delta_{21}+a^m}{2}L^m\right)
  \label{36}
\end{eqnarray}
into (\ref{34}), we obtain
\begin{eqnarray}
  S_{\mu e}^{\ell} =
-i \exp\left(-i \frac{\Delta_{21}L+2a^mL^m+a^cL^c}{2}\right)
F(\phi^m_{\ell},\phi^c_{\ell};\theta^m_{\ell},\theta^c_{\ell}),
\label{37}
\end{eqnarray}
where the function $F$ is defined by
\begin{eqnarray}
F(\phi^m,\phi^c;\theta^m,\theta^c)
=\sin 2 \phi^m \cos \phi^c\sin 2 \theta^m
+\cos^2 \phi^m \sin \phi^c\sin 2 \theta^c
+\sin^2 \phi^m \sin \phi^c\sin (2 \theta^c-4 \theta^m).
    \label{38}
 \end{eqnarray}
We can easily extract the physical meaning from this expression,
although this function $F$ becomes the same one as given in refs.
\cite{Petcov9903, Akhmedov9805} after a short calculation.
We describe the meaning of each term later.

Next, we calculate the amplitude $S_{\tau e}^h$
taking the limit $\alpha \to 0$.
In this limit, $S^m_{\mu e}$, $S^m_{\mu\tau}$ and so on vanish,
so the amplitude $S_{\tau e}^h$ in three layer
is calculated as
\begin{eqnarray}
S_{\tau e}^{h}
&=& S^m_{\tau e}S^c_{ee}S^m_{ee}
+ S^m_{\tau\tau}S^c_{\tau e}S^m_{ee}
+ S^m_{\tau e}S^c_{e\tau}S^m_{\tau e}
+ S^m_{\tau\tau}S^c_{\tau\tau}S^m_{\tau e}.
\label{39}
\end{eqnarray}
Substituting (\ref{16}) and
\begin{eqnarray}
\label{40}
&&  S^{m}_{ee}=\left[\exp (-iH_h^m L) \right]_{ee}
=\left(\cos \phi_h^m+i\cos 2 \theta_{h}^m
\sin \phi_h^m\right)
\exp\left(-i\frac{\Delta_{31}+a^m}{2}L^m\right),
\\
&&  S^{m}_{\tau\tau}=\left[\exp (-iH_h^m L^m) \right]_{\tau\tau}
=\left(\cos \phi_h^m-i\cos 2 \theta^{m}_{h}
\sin \phi_h^m\right)
\exp\left(-i\frac{\Delta_{31}+a^m}{2}L^m\right)
\label{41}
\end{eqnarray}
into (\ref{39}), we obtain
\begin{eqnarray}
  S_{\tau e}^h =
-i \exp\left(-i \frac{\Delta_{31}L+2a^mL^m+a^cL^c}{2}\right)
F(\phi^m_h,\phi^c_{h};\theta^m_{h},\theta^c_{h}),
\label{42}
\end{eqnarray}
which corresponds to equation (\ref{37}) by replacing the subscript and
superscript as $({\ell}) \to ({h})$.

Substituting (\ref{37}) and (\ref{42}) into (\ref{9})-(\ref{11}),
the coefficients $A$, $B$ and $C$ in three layers are given by
\begin{eqnarray}
  A\simeq&&
\hspace{-1.5em}
\sin 2\theta_{23} \cos \left(\frac{\Delta_{32}L}{2}\right)
F(\phi^m_{\ell},\phi^c_{\ell};\theta^m_{\ell},
\theta^c_{\ell})
F(\phi^m_h,\phi^c_{h};\theta^m_{h},
\theta^c_{h}),
  \label{43}
 \\
  B\simeq&&\hspace{-1.5em}
\sin 2\theta_{23} \sin \left(\frac{\Delta_{32}L}{2}\right)
F(\phi^m_{\ell},\phi^c_{\ell};\theta^m_{\ell},
\theta^c_{\ell})
F(\phi^m_h,\phi^c_{h};\theta^m_{h},
\theta^c_{h}),
  \label{44}
 \\
  C\simeq&&\hspace{-1.5em}
c_{23}^2
F(\phi^m_{\ell},\phi^c_{\ell};\theta^m_{\ell},
\theta^c_{\ell})^2
+s_{23}^2 F(\phi^m_h,\phi^c_{h};\theta^m_{h},
\theta^c_{h})^2.
  \label{45}
\end{eqnarray}
Thus, we can calculate the coefficients $A$ and $B$,
which are related to the magnitude of the CP effect,
by using our approximate formula.
We can see the following from the expressions of
$A$, $B$ and $C$.
The expression of $C$ is given as the sum of $F_h$ and
$F_{\ell}$, where we use the abbreviation $F_h$ and $F_{\ell}$ as
$F(\phi^m_h,\phi^c_{h};\theta^m_{h},\theta^c_{h})$ and
$F(\phi^m_{\ell},\phi^c_{\ell};\theta^m_{\ell},\theta^c_{\ell})$.
On the other hand, the expressions $A$ and $B$ are both given as
the product of $F_h$ and $F_{\ell}$ and furthermore multiplied by
the oscillating factor related to $\Delta_{32}$.
This is the main difference between $A$, $B$ and $C$.
However, all the coefficients depend on the function $F$.
In the following, we study the behavior of this function $F$.

At first, we divide $F$ given in (\ref{38}) into three parts as
\begin{eqnarray}
F(\phi^m,\phi^c;\theta^m,
\theta^c)&=&F_1+F_2+F_3, \label{46} \\
F_1&=&\sin 2 \phi^m \cos \phi^c\sin 2 \theta^m, \label{47} \\
F_2&=&\cos^2 \phi^m \sin \phi^c\sin 2 \theta^c, \label{48} \\
F_3&=&\sin^2 \phi^m \sin \phi^c\sin (2 \theta^c-4 \theta^m).
    \label{49}
 \end{eqnarray}
This separation of the function $F$ is useful to understand,
which contribution becomes large in the amplitude
because $F_1$, $F_2$ and $F_3$ correspond to the MSW effect
in the mantle, and in the core, and the mantle-core effect,
respectively.
By using the above expressions, the following interpretation in refs.
\cite{Petcov9903, Akhmedov9805} can be understood more clearly.

\begin{enumerate}
\item
$\cos 2 \phi^m=0$ and $\sin \phi^c=0$ \\
Only $F_1$ remains and the function takes the form
$F=\pm\sin 2 \theta^m$ because $F_2=F_3=0$ due to $\sin \phi^c=0$.
In the case that the above conditions are approximately satisfied
around the MSW resonance region of the mantle,
namely around the energy determined by $\sin 2 \theta^m=\pm 1$,
the function $F$ is enhanced.

\item
$\sin \phi^m=0$ and $\cos \phi^c=0$ \\
Only $F_2$ remains and the function takes the form
$F=\pm\sin 2 \theta^c$ because $F_1=F_3=0$ due to
$\sin \phi^m=0$ and $\cos \phi^c=0$.
In the case that the above conditions are approximately satisfied
around the MSW resonance region of the core,
namely around the energy determined by $\sin 2 \theta^c=\pm 1$,
the function $F$ is enhanced.

\item
$\cos \phi^m=0$ and $\cos \phi^c=0$ \\
Only $F_3$ remains and the function takes the form
$F=\pm \sin (2 \theta^c-4 \theta^m)$ because $F_1=F_2=0$
due to $\cos \phi^m=0$ and $\cos \phi^c=0$.
Around the energy determined by $\sin (2 \theta^c-4 \theta^m)=\pm 1$,
the function $F$ is enhanced.
This can be large, even if both effective mixing angles in the mantle
and in the core, $\theta^m$ and $\theta^c$, are small.
It is considered as the mantle-core effect.
It is realized in the case that $\Delta_{31}$ takes
the intermediate value of the matter potentials $a^m$ and $a^c$,
respectively, for the mantle and the core.
\end{enumerate}

\subsection{Interpretation of Numerical Results}

\hspace*{\parindent}
In this subsection, the numerical result
for $L=12000$km can be understood,
by using the analytical expression derived in the previous subsection.
All the coefficients $A, B$ and $C$
are determined by the functions $F_{\ell}$ and  $F_h$.
Here, we study the behavior of $F_{\ell}$ and $F_h$ in
the energy region larger than $E=1$ GeV.
We can approximate $F_{\ell}$ by using the fact $\Delta_{21}\ll a$
at $E>1$ GeV.
That is, the oscillation part and the mixing angle are
approximated by
\begin{eqnarray}
\phi_{\ell}&=&\frac{\Delta_{\ell} L}{2}\simeq \frac{aL}{2}
\sim {\rm const.}
\label{50}
 \\
\sin 2 \theta_{\ell}&=&
\frac{\Delta_{21}\sin 2 \theta_{12}^{}
}{\sqrt{(\Delta_{21}\cos 2 \theta_{12}-a)^2
+\Delta_{21}^2\sin^2 2 \theta_{12}}}
\simeq \frac{\Delta_{21}\sin 2 \theta_{12}}{a}
\propto \frac{1}{E}
\label{51}
\end{eqnarray}
from (\ref{15}).
As a result, we can also approximate $F_{\ell}$ from
(\ref{46})-(\ref{49}) as
\begin{eqnarray}
F_{\ell}
\propto \frac{1}{E}.
    \label{52}
 \end{eqnarray}
Thus, the value of $F_{\ell}$ decreases proportional to
the inverse of the neutrino energy.

On the other hand, some of the peaks appear in $F_h$
corresponding to $F_{h1}$, $F_{h2}$ and $F_{h3}$,
since $F_h$ includes the 1-3 MSW effect in the considered energy range.
Fig. 5 shows the component of $F_h$ by using our analytical
expression (\ref{46}),
where we use the matter densities in the mantle and the core
as $\rho^m=4.7$g/cm$^3$ and $\rho^c=11.0$g/cm$^3$,
the electron fraction as $Y_e^m=0.494$ and $Y_e^c=0.466$,
calculated by the PREM in the case of the baseline $L=12000$ km.

\begin{center}
\hspace{-1cm}
\begin{tabular}{cc}
    Component of $F_h^2$ ($\sin^2 2\theta_{13}=0.10$)& \quad
    Component of $F_h^2$ ($\sin^2 2\theta_{13}=0.04$)\\
\vspace{-0.3cm}

\hspace{-0.3cm}
    \resizebox{80mm}{!}{\includegraphics{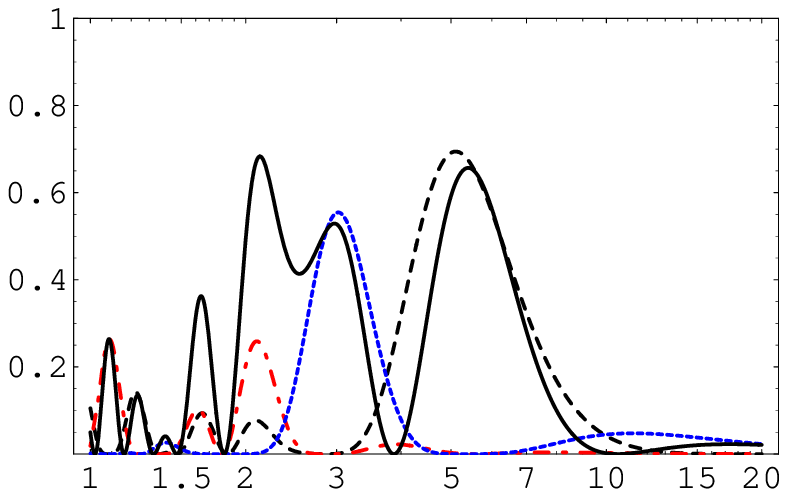}} &
    \resizebox{80mm}{!}{\includegraphics{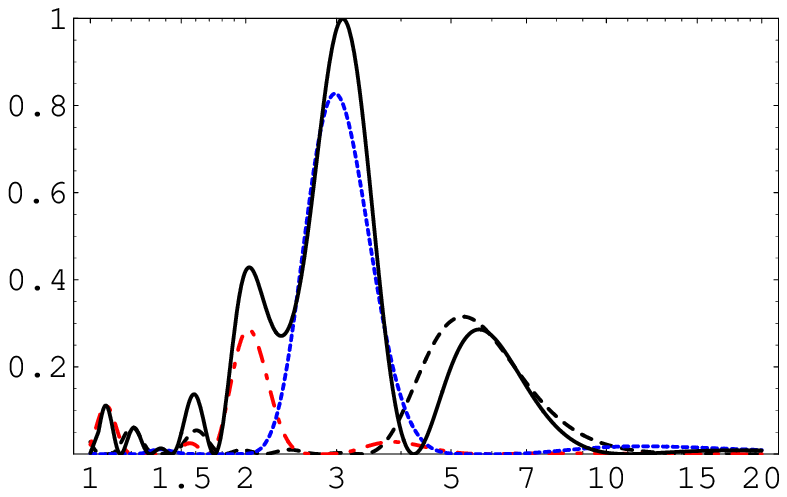}} \\
\vspace{0.3cm}
    \qquad E (GeV) & \qquad E (GeV) \\
\end{tabular}

\begin{flushleft}
Fig. 5 Components of $F_h^2$ calculated from our analytical formula with
$\sin^2 2\theta_{13}=0.10$ and $0.04$ from left to right.
Solid, dashed, dash-dotted and dotted lines correspond to
$F_h^2$, $F_{h1}^2$,
$F_{h2}^2$ and $F_{h3}^2$, respectively.
\end{flushleft}
\end{center}

In these figures, the solid line shows the magnitude of $F_h^2$,
and the dashed, dash-dotted and dotted lines show
the magnitude of $F_{h1}^2$, $F_{h2}^2$ and $F_{h3}^2$, respectively.
Furthermore, we represent the values of each component
$F_{h1}$, $F_{h2}$ and $F_{h3}$ at three peaks in Table 2.
\begin{center}
\begin{tabular}{cccccc} \hline
$\sin^2 2\theta_{13}$ & $E$ (GeV) & $F_h$ & $F_{h1}$ & $F_{h2}$ & $F_{h3}$
 \\ \hline
$0.10$ & $2.1$ & $0.821$ & $0.278$ & {\boldmath $0.509$} & $0.034$ \\
$0.10$ & $3.0$ & $0.726$ & $-0.029$ & $0.011$ & {\boldmath $0.744$} \\
$0.10$ & $5.4$ & $-0.810$ & {\boldmath $-0.821$} & $0.015$ & $-0.004$ \\
\hline
$0.04$ & $2.0$ & $0.648$ & $0.093$ & {\boldmath $0.532$} & $0.023$ \\
$0.04$ & $3.1$ & $0.999$ & $0.051$ & $0.055$ & {\boldmath $0.893$} \\
$0.04$ & $5.7$ & $-0.535$ & {\boldmath $-0.545$} & $0.013$ & $-0.003$ \\
\hline
\end{tabular}
\end{center}
\begin{flushleft}
Table 2. Components of $F_h$ calculated from our analytical formula with
$\sin^2 2\theta_{13}=0.10$, and $0.04$.
\end{flushleft}

Fig. 5 and Table 2 show that the peak in the right hand side
is dominated by $F_{h1}$ and mainly depends on the MSW effect
in the mantle.
The MSW resonance in the mantle is realized at the condition
$a^m=\Delta_{31}\cos 2 \theta_{13}$.
The energy determined by this condition is
$E\sim \frac{\Delta_{31}\cos 2 \theta_{13}}{2\sqrt{2}GN_e^m}\sim 5.7$ GeV.
The peak in the left hand side is dominated by $F_{h2}$ and
mainly depends on the MSW effect in the core.
The MSW resonance in the core is realized at the condition
$a^c=\Delta_{31}\cos 2 \theta_{13}$.
Noticing the relation $a^c\simeq 2.5\times a^m$,
the peak energy is given by around
$E\sim 5.7/2.5\sim 2.3$ GeV.
The energy of these peaks do not largely depend on the value $\theta_{13}$
in the case of $\sin^2 2\theta_{13}\ll 1$.
Furthermore, it is shown that
the mantle-core effect
mainly contributes to the peak at the center, when $F_{h3}$ becomes large.
The energy determined by the condition
$\sin(2\theta^c-4 \theta^m)\sim 1$ is about $E=3$-$4$ GeV
for $\sin^2 2\theta_{13}=0.04$.
In the case of $\sin^2 2\theta_{13}=0.10$, this condition cannot be
satisfied in any energy region and as a result the enhancement is weakened.
This phenomena is interesting because the
value of $F_h$ for $\sin^2 2\theta_{13}=0.04$ (small mixing) is
larger than that for $\sin^2 2\theta_{13}=0.10$ (large mixing).
It is interpreted as the total neutrino conversion pointed out
by Petcov {\it et al.} \cite{Petcov9903}.

Next, let us study how we can understand the behavior
of $A$, $B$ and $C$.
>From (\ref{49}), $C$ is approximated by
\begin{eqnarray}
C=\frac{1}{2}(F_{\ell}^2+F_h^2) \simeq \frac{1}{2}F_h^2,
\label{53}
\end{eqnarray}
where we neglect $F_{\ell}$, because of its smallness
compared to $F_h$ as shown in fig. 5.
Actually, the $C$-function has almost half of the size
of the $F_h^2$-function.
Therefore, $C$ has three peaks as $F_h^2$.
$P(\nu_e \to \nu_{\mu})$ in
ref. \cite{Akhmedov9808} corresponds to $C$ in this paper.
It means that the terms $A$ and $B$, related to the CP phase, were not
considered in previous papers.

Next, we obtain the expressions for $A$ and $B$ from
(\ref{47}) and (\ref{48}) as
\begin{eqnarray}
A&\simeq&
\cos \left(\frac{\Delta_{32}L}{2}\right)F_{\ell}F_{h}
\propto \frac{1}{E}
\cos \left(\frac{\Delta_{32}L}{2}\right)F_{h},
  \label{54}
 \\
B&\simeq&
\sin \left(\frac{\Delta_{32}L}{2}\right)F_{\ell}F_{h}
\propto \frac{1}{E}
\sin \left(\frac{\Delta_{32}L}{2}\right)F_{h}.
  \label{55}
\end{eqnarray}
>From these expressions, we can see the following.
First, the mantle-core effect, which is different from the usual MSW
effect, appears not only in $C$ but also in $A$ and $B$ because
of the multiplication of $F_h$.
Second, $A$ and $B$ are suppressed compared
with $C$ in the energy range $E>1$ GeV because of $F_{\ell}\propto 1/E$.
Third, $A$ and $B$ depend on
the oscillation part
$\frac{\Delta_{32}L}{2}$ additionally to $L/E$
dependence included in $F$.
Because of this factor,
the oscillation phases of $A$ and $B$ have a difference of
about a quarter of the wavelength.

\section{Summary}
 \label{sec:summary}

\hspace*{\parindent}
In this paper, we investigate the matter effect included in
the terms related to the CP phase, particularly
in the case that neutrinos pass through the earth core.
The results are summarized as follows.
\begin{enumerate}
\item
Our approximate formulas (\ref{2})-(\ref{4}) include
non-perturbative effect of the small parameters
$\alpha=\Delta_{21}/\Delta_{31}$ and $s_{13}$.
As a result, the precision of the formula is rather improved
compared to the formula which includes up to second order of
$\alpha$ and $s_{13}$ around the MSW resonance regions.

\item
We numerically calculate the coefficients $A$, $B$ and $C$ for the
baseline length $L=6000$ km and $12000$ km
by using the PREM as the earth matter density.
As a result, the magnitude of $A$ and $B$ can reach
a few ten \% of $C$ around the three main peaks of $C$
even for the case of including non-perturbative effect.

\item
We give the qualitative understanding of
the behavior for $A$, $B$ and $C$ by using
our approximate formula.
The mantle-core effect, which is different from the usual MSW
effect, appears not only in $C$ but also in $A$ and $B$,
although the effect is weakened.

\end{enumerate}

>From the results of this paper, it has been found that
the effects of the leptonic CP phase can be comparatively large
in the oscillation probability, when neutrinos pass through
the earth.
We should consider the CP phase effects more seriously
in order to extract the information on $\theta_{13}$ and
the sign of $\Delta m_{31}^2$ in future experiments.

\vspace{20pt}
\noindent
{\Large {\bf Acknowledgment}}

\noindent
We would like to thank Prof. Wilfried Wunderlich for helpful
comments and advice on English expressions.

\end{document}